\documentclass[review,5p,twocolumn,10pt]{elsarticle}

\usepackage[colorlinks,citecolor=red]{hyperref}  
\usepackage{threeparttable}  
\usepackage{booktabs}  

\usepackage{array}  

\usepackage[american]{babel}
\usepackage{microtype}
\usepackage{color}

\usepackage{graphicx}
\usepackage{caption}
\usepackage{graphicx}
\usepackage{dcolumn}
\usepackage{bm}
\usepackage{latexsym}
\usepackage{amsmath}
\usepackage{amsmath,amssymb,bm}
\usepackage{graphicx,color}
\usepackage{hyperref}
\usepackage{epstopdf}
\usepackage{mathrsfs}
\usepackage{float}
\usepackage{subfigure}
\usepackage{braket}

\newcommand{\be}{\begin{equation}}
\newcommand{\ee}{\end{equation}}

\usepackage{amssymb}
\usepackage{amsthm}
\usepackage{amsmath}

\usepackage{lineno}
\journal{Entropy}

\begin{document}


\begin{frontmatter}



\title{\textbf{ Anomalous Coulomb-Enhanced Charge Transport in Triangular Triple Quantum Dots Systems }}


\author{Shuo Dong $^{a}$}

\author{Junqing Li $^{a}$}

\author{Jianhua Wei $^{a*}$}
\ead{corresponding author: wjh@ruc.edu.cn}


\address[1]{School of Physics, The Renmin University of China, Beijing, 100876 China}

\begin{abstract}
Electron correlation and quantum interference are pivotal in mesoscopic transport. We theoretically study the nonequilibrium transport dynamics of a triangular triple quantum dot (TTQD) molecule connected to fermionic reservoirs using the exact hierarchical equations of motion (HEOM)
formalism. We demonstrate a counter-intuitive transport signature where
the stationary current is significantly enhanced by increasing the $U$, a behavior distinct from the
suppression typically observed in linear quantum dot arrays. By analyzing
the evolution of spectral functions, we attribute this enhancement to
the interplay between Coulomb interaction-induced energy shifts and quantum interference effects unique to the triangular topology. We also explore how the circulation of chiral currents and electrode coupling strength
modulates these interaction effects. Finally, we present a three-dimensional map of the
transport current as a function of inter-dot tunneling ($t$) and Coulomb
interaction ($U$), illustrating their combined effect on the current
magnitude and its applications.
\end{abstract}

\end{frontmatter}

\section{Introduction}
Quantum dots (QDs), often described as ``artificial atoms,'' have become one of the most widely studied platforms in condensed matter physics and nanoscience. These nanoscale semiconductor structures confine electrons in all three spatial dimensions, yielding discrete, atom-like energy levels that are tunable through external gate voltages and magnetic fields~\cite{1,2}. This tunability has made quantum dots a natural setting for studying fundamental quantum phenomena in controllable mesoscopic systems and for developing quantum technologies. The foundation of quantum dot physics was established through the experimental observation of single-electron charging effects. Meirav, Kastner, and Wind demonstrated periodic conductance resonances in GaAs nanostructures, providing the first clear evidence of Coulomb blockade in quantum
dots~\cite{3}. These pioneering experiments showed that when a quantum dot is weakly
coupled to electron reservoirs, the addition energy required to place each successive electron
on the dot produces a distinctive pattern of conductance peaks as a function of gate voltage.
Kouwenhoven and colleagues further advanced the understanding of discrete energy levels and
Coulomb charging effects, establishing quantum dots as true artificial atoms with shell
structure~\cite{4,5}.

Central to quantum dot physics is the on-site Coulomb repulsion energy $U$, which
characterizes the electrostatic cost of placing two electrons on the same dot. The theoretical framework for understanding $U$ was established by Beenakker,
who provided an early treatment of Coulomb blockade oscillations in quantum dot
conductance~\cite{6}. In the Hubbard model description of quantum dots~\cite{7}, $U$
competes directly with the kinetic energy scale set by the interdot tunneling amplitude $t$.
When $U \ll t$, electrons delocalize freely and transport is relatively unimpeded. As $U$
increases toward and beyond $t$, charge fluctuations become energetically costly, electrons
tend to localize on individual dots, and the system enters a strongly correlated regime in
which conventional single-particle descriptions break down. This competition between $U$ and
$t$ is a genuine control parameter that drives the system through qualitatively different
physical regimes, from weakly interacting Fermi liquid behavior to strongly correlated
Mott-like insulating states. How $U$ shapes transport properties is thus a central issue in
quantum dot physics. Loss and DiVincenzo's seminal proposal established electron spins
confined in quantum dots as viable qubits, highlighting their long coherence times and
gate-voltage controllability as key advantages for scalable quantum computation~\cite{8}.
Following this proposal, subsequent experiments demonstrated precise control over individual
and coupled spin qubits~\cite{9,10,11,12}. These developments established quantum dots as
leading candidates for solid-state quantum computing, and simultaneously underscored the
importance of understanding Coulomb interactions, since $U$ directly governs the exchange
splitting between singlet and triplet spin states that underpins spin qubit operation.

Linear quantum dot arrays, from double quantum dots (DQDs) to triple quantum dots
(TQDs) systems, have been extensively studied as building blocks for quantum
computation and as platforms for exploring interdot coupling, tunneling dynamics,
and multi-qubit operations. For DQDs, Van der Wiel et al.\ provided a
comprehensive review of electron transport, discussing the interplay among
interdot tunneling, lead coupling, and Coulomb interactions~\cite{13}; coherent
charge oscillations were demonstrated by Hayashi et al.~\cite{14}, and the Pauli
spin blockade was observed by Ono et al.\ and Johnson et al.~\cite{15,16}.
Extending to TQDs, Schr\"oer et al.\ realized electrostatically defined serial
devices with controllable electron numbers~\cite{19}, while Gaudreau and
colleagues fabricated tunable few-electron TQDs and observed coherent three-dot
coupling~\cite{20,21}; the transport theory was further developed by Rogge and
Haug, who highlighted quantum interference between different tunneling
pathways~\cite{22,23}, and by Michaelis et al.\ using rate equations~\cite{24}.
Across both DQD and linear TQD configurations, the role of the on-site Coulomb
interaction $U$ is qualitatively consistent: stronger repulsion raises the
addition energy, widens the Coulomb blockade valleys, localizes electrons on
individual dots, and suppresses sequential tunneling, reinforcing the
conventional expectation that $U$ impedes rather than facilitates current flow.

Moving beyond linear quantum dots, triangular triple quantum dots(TTQDs) systems are particularly intriguing because they break linear symmetry and
introduce geometric frustration, potentially giving rise to novel quantum phenomena. The
threefold rotational symmetry of triangQDs creates fundamentally different interference
conditions compared to linear arrangements. Korkusinski et al.\ investigated the electronic
properties of triangular lateral quantum dot molecules and introduced ``topological Hund's
rules'' to describe state filling~\cite{25}, showing that geometry alone can qualitatively
alter the energy spectrum and charge configurations. Experimentally, Granger et al.\ mapped
out the three-dimensional transport diagram of a TQD, revealing complex charge stability
regions in multi parameter space~\cite{26}. Seo et al.\ demonstrated charge
frustration effects in a triangular TQD arising from the geometric arrangement~\cite{27},
showing that the triangular geometry leads to degenerate charge configurations and
non-trivial ground states that differ qualitatively from linear arrangements. On the
theoretical side, \v{Z}itko, Bon\v{c}a, and Pruschke investigated quantum phase transitions
in TQDs using numerical renormalization group methods, finding that the interplay between
Kondo screening and geometric structure can lead to complex phase diagrams~\cite{28}.

Yet despite this extensive body of work, a basic question about transport in triangular TQDs
has not been examined systematically. For single dots and linear multi-dot systems, it is
well established that increasing by making charge fluctuations
energetically costly and driving the system toward electron localization\cite{29,30,31}. However, this picture was developed
primarily for systems with simple, effectively one-dimensional tunneling topologies, in which
a single dominant transport pathway exists and $U$ acts uniformly to block it. The
triangular TQD geometry violates this assumption in a fundamental way. With three dots
arranged in a closed loop, electrons have access to multiple interfering paths whose phase
relationships are not fixed by geometry alone but are renormalized by interactions. The
closed-loop structure introduces a topological character analogous to Aharonov-Bohm physics
even in the absence of external magnetic fields. Geometric frustration prevents the
simultaneous optimization of all pairwise interactions, potentially stabilizing unusual
ground states. The absence of left-right mirror symmetry distinguishes the triangular
configuration from all linear arrangements and removes constraints that would otherwise pin
the interference conditions. These features together raise a question that has not been
addressed: does increasing $U$ suppress conductance in a triangular TQD as in simries, or can the interplay between strong Coulomb correlations and closed-loop
topology produce qualitatively different behavior?

Addressing this question matters both for fundamental reasons and for practical ones.
Identifying a regime in which strong correlations enhance rather than suppress transport
would pointopic physics, one that expands our understanding of
how electron-electron interactions operate in confined geometries. At the same time, quantum
dot arrays are actively being developed for quantum simulation, computation, and sensing,
and triangular geometries are increasingly explored as building blocks for two-dimensional
quantum dot arrays. If $U$ can enhance conductance in such geometries, this opens a design
principle for interaction-controlled quantum devices. For TQD-based qubits specifically,
understanding how $U$ affects transport and coherence across a wide parameter range is
essential for device optimization, since qubit operation requires precise control over both
charge and spin degrees of freedom that $U$ directly governs.

In this work, we investigate the transport properties of a triangular TQD across a wide
range of Coulomb interaction strengths using  hierarchical equations of motion (HEOM) method. Our
central finding is that increasing the on-site Coulomb repulsion $U$ enhances conductance
through the system, in stark contrast to the conventional expectation that stronger
interactions suppress current flow. This effect is intrinsically
tied to the closed-loop triangular geometry and does not arise in linear TQD configurations,
confirming that the interplay between $U$ and geometric structure drives the phenomenon. Our
results demonstrate a mechanism by which strong correlations can be harnessed
constructively, with potential implications for interaction-controlled quantum devices and
for the broader understanding of correlated transport in geometrically frustrated systems.

\section{Theory and Methoed}

The theoretical treatment of transport through interacting quantum dot systems in the
nonequilibrium regime requires sophisticated many-body techniques.

The system consists of three quantum dots arranged in a triangular geometry, as shown in figure 1(a), where all on-site energies and intra-dot Coulomb interactions are identical, and the inter-dot hopping amplitude $t$ is uniform across all three links. Two of the three quantum dots (labeled dot 1 and dot 3) are each coupled to a separate fermionic electron reservoir, and a small bias voltage $V$ is applied between the two reservoirs to drive a transport current through the system. Another quantum dot (dot 2) is not directly coupled to any reservoir but participates in transport indirectly through its hybridization with dots 1 and 3. Throughout the calculation, electron-hole symmetry is maintained by setting the on-site energy $\varepsilon_d = -U/2$, where $U$ is the Coulomb interaction strength. 

The Anderson impurity model provides the theoretical framework for studying this system, in the multi-impurity generalization, each quantum dot is characterized by an on-site energy level, an intra-dot Coulomb repulsion $U$, and hybridization both with neighboring dots through hopping $t$ and with the continuum of conduction electrons in the attached reservoirs. The Hamiltonian of the full system can be written as
\begin{equation}
H = H_{\mathrm{dots}} + H_{\mathrm{leads}} + H_{\mathrm{hyb}},
\end{equation}
where $H_{\mathrm{dots}}$ contains the on-site energies, the Coulomb repulsion, and the inter-dot hopping among all three dots:
\begin{equation}
		H_{\mathrm{dots}}=\sum_{i \mu} \varepsilon_{i} \widehat{d}_{i \mu}^{\dagger} \widehat{d}_{i \mu}+U \sum_{i} \widehat{n}_{i \uparrow} \widehat{n}_{i \downarrow}+\sum_{i \neq j \mu} t_{i j} \widehat{d}_{i \mu}^{\dagger} \widehat{d}_{j \mu}.
	\end{equation} 
The $d_{i \mu}^{\dagger}\left(d_{i \mu}\right)  $ in above fomula is creation(annihilation) operator for an electron with $\mu $ spin on the $i-th $ dot.$H_{\mathrm{leads}}$ describes the two noninteracting electron reservoirs:
\begin{equation}
		H_{\mathrm{leads}}=\sum_{\alpha k \mu} \varepsilon_{\alpha k} c_{\alpha k \mu}^{\dagger} c_{\alpha k \mu} ,
	\end{equation}
	$ c_{\alpha k \mu}^{\dagger}\left(c_{\alpha k \mu}\right)$ is creation(annihilation) operator for electron of lead $ \alpha$ on the $ k -th $ state, and $\epsilon_{k \alpha}$  is the energy of an electron with wave vector  $k $ in the $ \alpha$  lead. And $H_{\mathrm{hyb}}$ captures the tunnel coupling between dots 1 and 3 and their respective leads:
\begin{equation}
		H_{\mathrm{hyb}}=\sum_{\alpha k i \mu} V_{\alpha k i \mu} d_{i \mu}^{\dagger} c_{\alpha k \mu}+  H.c. ,
	\end{equation}
	
with $V_{\alpha k i \mu \mu} $ being the tunnel matrix element between  $i -th$ impurity and electrons with $ k -th$ state on the  $\alpha$ -reservoir. For this paper, $ V_{\alpha k \mu} $ is the electron tunneling strength between two leads. The effect of electron reservoirs on QDs is taken into account through the hybridization functions, $ \Delta_{\mu v}(\omega) \equiv \Sigma_{\alpha} \Delta_{\alpha \mu \nu}(\omega)=\pi \Sigma_{\alpha k} V_{\alpha \mu k} V_{\alpha v k}^{*} \delta\left(\omega-\varepsilon_{\alpha k}\right) $, in the absence of applied chemical potentials. Generally, we adopt Lorentzian hybridization functions in the HEOM approach, that is, $ \Delta_{\mu v}(\omega)=\delta_{\mu v} \Delta W^{2} /\left(\omega^{2}+W^{2}\right) $, with $ \Delta=\Sigma_{\alpha} \Delta_{\alpha} $ being the overall dot-lead coupling strength and  W  is the bandwidth of the electrodes.

A magnetic flux $\phi$ threading the lead-free triangular loop modifies the inter-dot hopping amplitude through a phase factor. We use perturbation theory in the inter-dot tunneling $t$, treating it as small compared to the on-site Coulomb repulsion $U$. This approach allows us to derive an effective spin-exchange Hamiltonian from the microscopic quantum dot Hamiltonian $H_{\mathrm{dots}}$\cite{32,33} by integrating out high-energy charge excitations. The resulting Hamiltonian takes the form: 

	\begin{equation}
		\begin{aligned}
			H_{\text {eff }}= & -t(1-n) \sum_{j k, \mu}\left(\hat{d}_{j \mu}^{\dagger} \hat{d}_{k \mu}+\text { H.c. }\right) \\
			& +J \sum_{j<k}\left(\hat{\boldsymbol{S}_{j}} \hat{\boldsymbol{S}_{k}}-\frac{1}{4} \hat{n}_{j} \hat{n}_{k}\right)+\chi \hat{\boldsymbol{S}_{1}}\left(\hat{\boldsymbol{S}_{2}} \times \hat{\boldsymbol{S}_{1}}\right),
		\end{aligned}
	\end{equation}

	where  $n$ is the the expectation value of the occupation number per dot, and $\hat{n}_{j}=\hat{d}_{j}^{\dagger} \hat{d}_{j}$ is the number operator for electrons on dot $j$. $\hat{\boldsymbol{S}_{1}}$, $\hat{\boldsymbol{S}_{2}}$ and $\hat{\boldsymbol{S}_{3}}$ are the spin operators on quantum dots, and $\mathbf{S}_{j}=\frac{1}{2} \sum_{\mu, \mu^{\prime}} \hat{d}_{j \mu}^{\dagger} \boldsymbol{\tau}_{\mu \mu^{\prime}} \hat{d}_{j \mu^{\prime}}$, $\boldsymbol{\tau}_{\mu \mu^{\prime}}$ are the Pauli matrices. The first term will vanish in the half-filling situation $(n=1)$. The second term is Heisenberg exchange interaction with $ J=   4 t^{2} / U$. The third term is the chiral term with chiral operator\cite{41} $ \widehat{S}_{1} \cdot\left(\widehat{S}_{2} \times \widehat{S}_{3}\right)$,  where $ \chi $ is the chiral interaction with  $\chi=24 t^{3} \sin \left(2 \pi \phi / \phi_{0}\right) / U^{2} $, and $ \phi $ is the magnetic flux enclosed by the TTQD structure. Here,  $\phi_{0}=h c / e $ is the unit of quantum flux. For simplicity, we let $ \varphi=2 \pi \phi / \phi_{0}$; thus, $ \chi=24 t^{3} \sin (\varphi) / U^{2}$.

The hierarchical equations of motion(HEOM) approach examines quantum dot properties in equilibrium and nonequilibrium states through the reduced density operator, which possesses a universal formalism applicable to arbitrary system Hamiltonians. It can accurately solves the three-impurity Anderson model\cite{34}. Recent progress has substantially broadened HEOM's scope and efficiency for quantum transport applications. Tanimura offered a comprehensive review of theoretical foundations and numerical implementations for both bosonic and fermionic systems, emphasizing its advantages over perturbative master equation approaches in strong-coupling regimes\cite{37}. These developments establish HEOM as a reliable tool for exploring thermoelectric phenomena in multi-dot systems.

The HEOM framework derives from Feynman-Vernon influence functional path-integral theory\cite{35} and employs Grassmann algebra for fermionic dissipation\cite{36}. Consequently, HEOM provides formally exact treatment of general open systems coupled to reservoir baths satisfying Grassmann Gaussian statistics. The mathematical construction has been comprehensively discussed in previous references.\cite{38,39,40} Here, we highlight key features particularly relevant to the quantum dot systems examined in this work's main text.

The HEOM approach provides a universal framework for investigating quantum dot properties under equilibrium and non-equilibrium conditions through the reduced density operator, applicable to arbitrary system Hamiltonians. We present a concise derivation below. At time $t$, the reduced density operator of the system is defined as $\rho(t)=\operatorname{tr}_{\text{res}} \rho_{\mathrm{T}}(t)$, where $\rho_{\mathrm{T}}(t)$ denotes the total density operator of both system and reservoirs. The reduced density operator $\rho(t)$ connects to its initial value at time $t_{0}$ through the reduced Liouville-space propagator $\mathcal{G}\left(t, t_{0}\right)$:

\begin{equation}
    \rho(t)=\mathcal{G}\left(t, t_{0}\right) \rho\left(t_{0}\right).
\end{equation}

Using the Feynman-Vernon influence functional formalism, the path-integral representation of the reduced Liouville-space propagator takes the form

\begin{equation}
    \mathcal{G}\left(\psi, t ; \psi_{0}, t_{0}\right)=\int_{\psi_{0}\left[t_{0}\right]}^{\psi[t]} \mathcal{D} \psi \mathrm{e}^{\mathrm{i} S[\psi]} \mathcal{F}[\psi] \mathrm{e}^{-\mathrm{i} S\left[\psi^{\prime}\right]},
\end{equation}

where $S[\psi]$ denotes the classical action of the reduced system and $\mathcal{F}[\psi]$ is the corresponding influence functional. Following the work of Zhenhua Li et al.\cite{42}, we invoke Wick's theorem together with Grassmann algebra to express the influence functional $\mathcal{F}[\psi]$ as

\begin{equation}
    \mathcal{F}[\psi]=\exp \left\{-\int_{0}^{t} \mathrm{~d} \tau \mathcal{R}[\tau,\{\psi\}]\right\},
\end{equation}

where $\mathcal{R}[\tau,\{\psi\}]=\frac{i}{\hbar^{2}} \sum_{\alpha i \mu \sigma} \mathcal{A}_{i j \mu}^{\bar{\sigma}}[\psi(t)] \mathcal{B}_{{\alpha} i \mu}^{\sigma}[t, \psi]$, with $\sigma=+,-$ and $\bar{\sigma}=-\sigma$. The Grassmann variables $\mathcal{A}_{i s}^{\bar{\sigma}}$ and $\mathcal{B}_{\alpha i \mu}^{\sigma}$ are defined as

\begin{equation}
    \mathcal{A}_{i s}^{\bar{\sigma}}[\psi(t)]=d_{i s}^{\sigma}[\psi(t)]+d_{i s}^{\sigma}\left[\psi^{\prime}(t)\right],
\end{equation}

\begin{equation}
    \mathcal{B}_{\alpha i \mu}^{\sigma}[t, \psi]=-\mathrm{i}\left[B_{\alpha i \mu}^{ \sigma}(t, \psi)-B_{\alpha i \mu}^{\prime \sigma}\left(t, \psi^{\prime}\right)\right],
\end{equation}

with

\begin{equation}
    \begin{array}{l}
        B_{\alpha i \mu}^{\sigma}(t, \psi)=\sum_{j} \int_{0}^{t} \mathrm{~d} \tau C_{\alpha i j \mu}^{\sigma}(t-\tau) d_{j \mu}^{\sigma}[\psi(\tau)], \\
        B_{\alpha i \mu}^{\prime \sigma}\left(t, \psi^{\prime}\right)=\sum_{j} \int_{0}^{t} \mathrm{~d} \tau C_{\alpha i j \mu}^{\bar{\sigma} *}(t-\tau) d_{j \mu}^{\sigma}\left[\psi^{\prime}(\tau)\right].
    \end{array}
\end{equation}

Here, $C_{\alpha i j \mu}^{\sigma}(t)$ represents the reservoir correlation functions. In the present computational scheme, $C_{\alpha i j \mu}^{\sigma}(t)$ is decomposed into a sum of exponential terms by combining the fluctuation-dissipation theorem with the Cauchy residue theorem and the Pad\'{e} spectral decomposition\cite{43} of the Fermi function:

\begin{equation}
    C_{\alpha i j \mu}^{\sigma}(t)=\sum_{m=1}^{M} \eta_{\alpha i j \mu m}^{\sigma} \mathrm{e}^{-\gamma_{\alpha i j \mu m}^{\sigma} t}.
\end{equation}

The bath effects enter the equations of motion through $M$ exponential terms. The auxiliary density operators (ADOs) $\left\{\rho_{j}^{n}=\rho_{j_{1} \ldots j_{n}}\right\}$ are determined by the time derivative of the influence functional. The composite index $j \equiv(\sigma \mu m)$ characterizes the transfer of an electron to or from ($\sigma=+/-$) the impurity state $\mu$, while $n=1,2,\ldots,L$ specifies the truncation tier level. In the present study, setting $L=4$ is found sufficient for obtaining numerically exact results. The resulting hierarchy equations of motion can be written in the following compact form:

\begin{equation}
    \begin{aligned}
        \dot{\rho}_{j_{1} \cdots j_{n}}^{(n)}= & -\left(\mathrm{i} \mathcal{L}+\sum_{r=1}^{n} \gamma_{j}\right) \rho_{j_{1} \cdots j_{n}}^{(n)}-\mathrm{i} \sum_{j} \mathcal{A}_{j} \rho_{j_{1} \cdots j_{n} j}^{(n+1)} \\
        & -\mathrm{i} \sum_{r=1}^{n}(-)^{n-r} \mathcal{C}_{j_{r}} \rho_{j_{1} \cdots j_{r-1} j_{r+1} \cdots j_{n}}^{(n-1)}.
    \end{aligned}
\end{equation}

In the above, $\mathcal{L}$ is the Liouvillian superoperator of the quantum dots, which incorporates electron-electron interactions, and is defined as $\mathcal{L} \cdot \equiv\left[H_{\mathrm{dot}}, \cdot\right]$. The Grassmannian superoperators $\mathcal{A}_{\bar{j}} \equiv \mathcal{A}_{i s}^{\bar{\sigma}}$ and $\mathcal{C}_{j} \equiv \mathcal{C}_{i j s m}^{\sigma}$ act on an arbitrary operator $\hat{O}$ according to $\mathcal{A}_{j} \hat{O} \equiv\left[\hat{d}_{i s}^{\hat{\sigma}}, \hat{O}\right]$ and $\mathcal{C}_{j} \hat{O} \equiv \eta_{j} \hat{d}_{i s}^{\sigma} \hat{O}+\eta_{j}^{*} \hat{O} \hat{d}_{i s}^{\sigma}$, respectively.

Furthermore, only the first-tier auxiliary density operators are required to accurately evaluate the transient current flowing through electrode $\alpha$:

\begin{equation}
    I_{\alpha}(t)=\mathrm{i} \sum_{i \mu} \operatorname{tr}_{\mathrm{sys}}\left[\rho_{\alpha i \mu}^{\dagger}(t) \hat{d}_{i \mu}-\hat{d}_{i \mu}^{\dagger} \rho_{\alpha i \mu}^{-}(t)\right].
\end{equation}

In addition, the spectral function is obtained by propagating the correlation functions $C_{a_{i \mu} a_{i \mu}^{\dagger}}(t)$ and $C_{a_{i \mu}^{\dagger} a_{i \mu}}(t)$ in time and subsequently performing a semi-infinite Fourier transform. This procedure directly yields the spectral function $A_{i \mu}(\omega)$ associated with spin $\mu$ in the $i$-th quantum dot. The fluctuation-dissipation theorem further allows the spectral density function of the quantum dots to be extracted from these correlation functions. The spectral function is given by

\begin{equation}
    A_{i \mu}(\omega)=\frac{1}{\pi} \operatorname{Re}\left\{\int_{0}^{\infty} d t\left\{C_{a_{i \mu} a_{i \mu}^{\dagger}}(t)+\left[C_{a_{i \mu}^{\dagger} a_{i \mu}}(t)\right]^{*}\right\} e^{-i \omega t}\right\}.
\end{equation}

\section{Results and Discussion}

Figure 1 presents a comparison of the transport current as a function of $U$ for the triangular triple quantum dots(TTQDs) and linear triple quantum dots(LTQDs) configurations, with all other parameters held identical and the inter-dot hopping fixed at $t = 0.25$ meV. In the linear triple quantum dot arrangement, where only nearest-neighbor hopping is present and the three dots form a chain, the transport current decreases monotonically as $U$ increases from $2t$ to $5t$. This monotonic suppression is consistent with the standard expectation from the Anderson impurity model in the strong-coupling regime. In striking contrast, the triangular triple quantum dot system exhibits a pronounced non-monotonic dependence of the transport current on $U$. Starting from $U = 2t$, the current first increases with $U$, reaches a maximum at approximately $U = 3.5t$, and subsequently decreases for larger values of $U$. This non-monotonic behavior represents the central finding of this work and signals the presence of a mechanism that, over a certain range of Coulomb interaction, enhances the transport current despite the growing on-site repulsion. 

\begin{figure}[ht] 
	\centering 
	\includegraphics[width=0.4\textwidth]{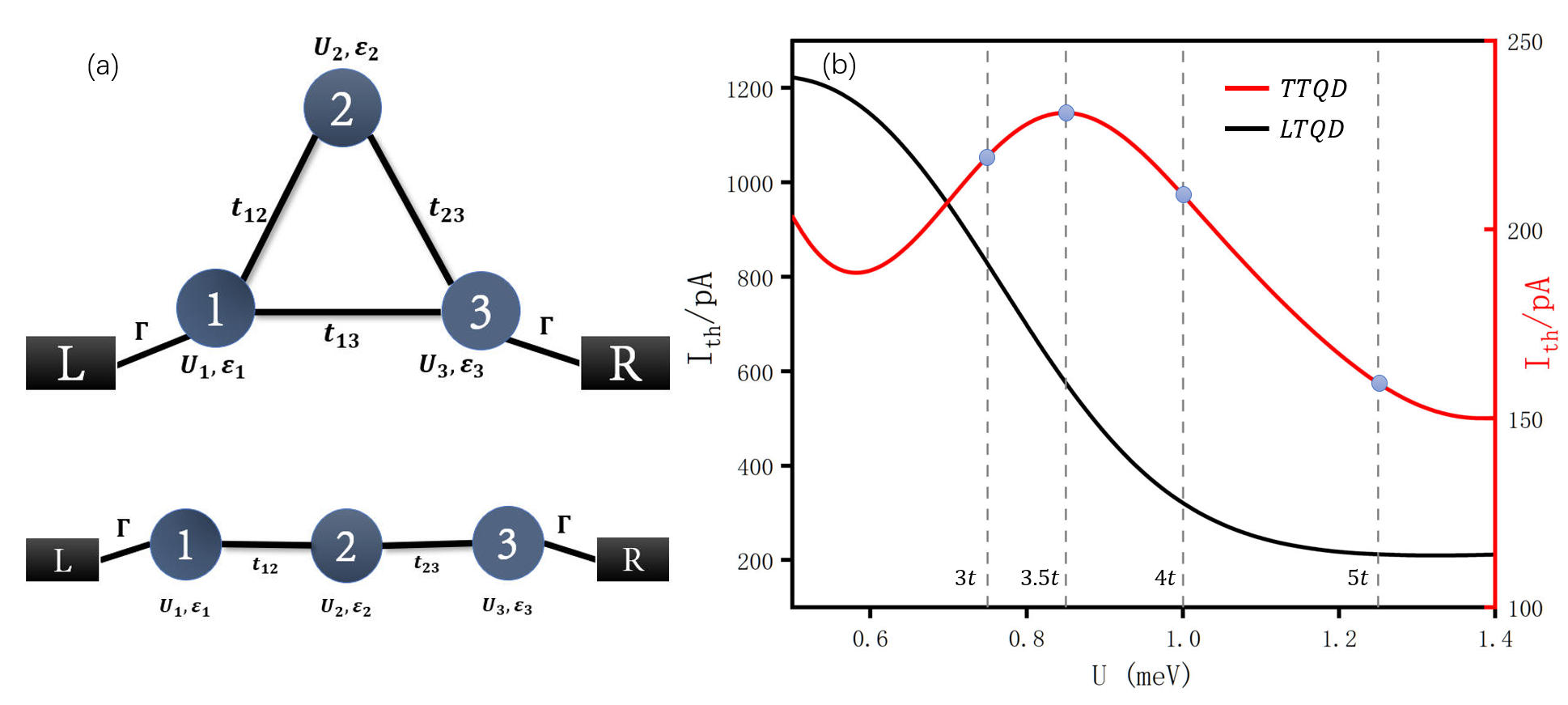} 
	\caption{(a)Schematic diagrams of a triangular triple quantum dots (TTQDs) and a linear triple quantum dots (LTQDs), Quantum dot 1 and quantum dot 3 are connected to the left and right electrodes.(b)Comparison of the transport current as a function of the on-site Coulomb interaction $U$ for the TTQDs and LTQDs configurations. The filled markers on the TTQD curve denote the specific values of $U$ at which the spectral function $A(\omega)$ has been explicitly evaluated, with all other parameters held identical and the inter-dot hopping fixed at $t = 0.25meV$ .The other parameters are set as follows:   $k_B T = 0.1meV$, $\Gamma =0.025meV$, $V=0.1meV$} 
	\label{fig:example-image} 
\end{figure}

To elucidate the mechanism underlying the non-monotonic current, we compute the frequency-resolved spectral function $A(\omega)$ at several representative values of $U$ along both the ascending and descending portions of the current curve, as shown in figure2. Within the small-bias regime, the current is predominantly controlled by the spectral weight available inside the conducting window around the Fermi level, which is shown by the dotted line in Figure 2. For clarity, we restrict our presentation to the spectral weight in the vicinity of the Fermi level $\omega = 0$.

\begin{figure}[ht] 
	\centering 
	\includegraphics[width=0.5\textwidth]{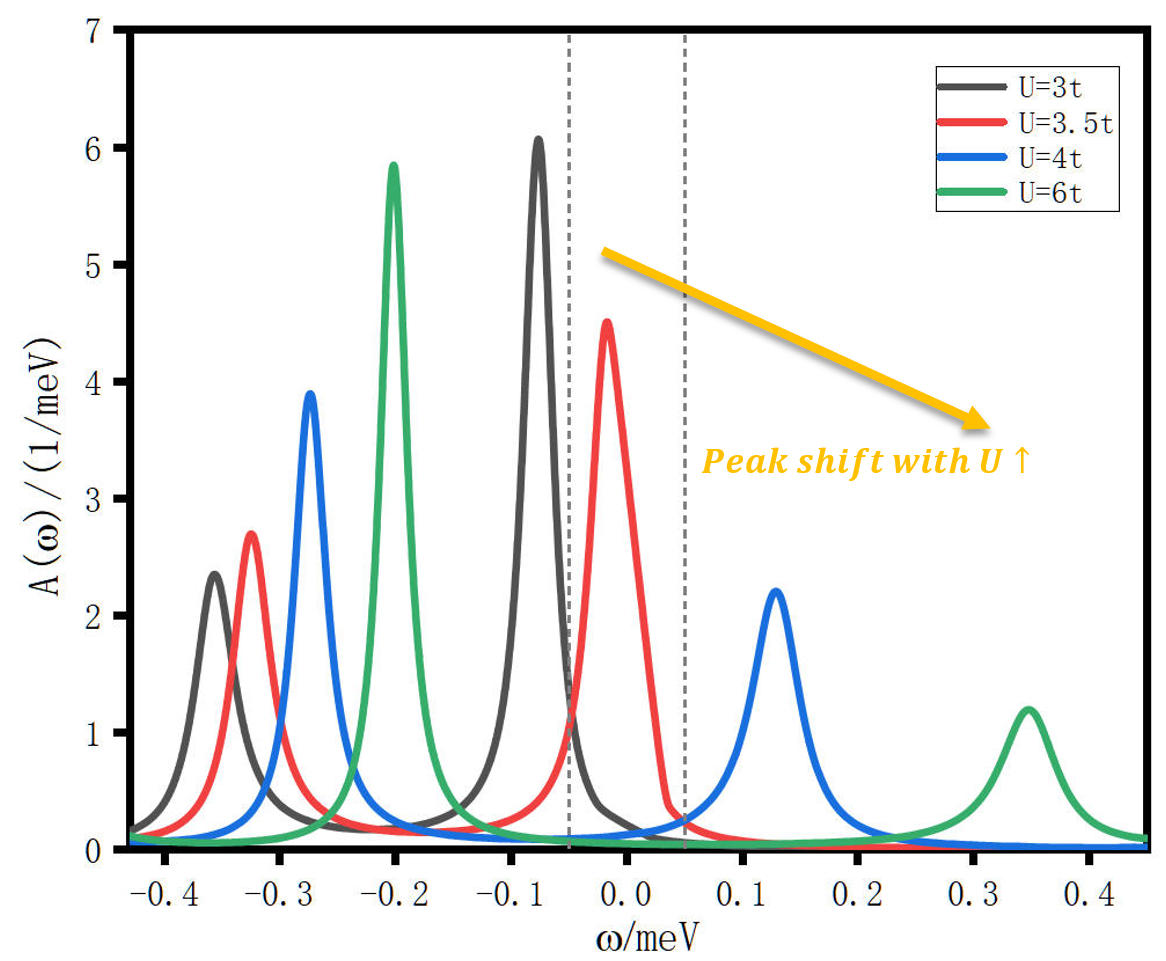} 
	\caption{Spectral function $A(\omega)$ of the triangular triple quantum dot system in the
    vicinity of the Fermi level $\omega = 0$, computed at representative
    on-site Coulomb interaction $U$ selected from both the rising and the falling
    portions of the current curve in Fig.~1.} 
	\label{fig:example-image} 
\end{figure}

At the smallest value of $U$ considered, both spectral peaks reside below the Fermi level. As $U$ increases beyond $2t$, a systematic evolution of the spectral function is observed. The right peak, being the one closest to the Fermi level, begins to shift toward the Fermi energy. Simultaneously, the spectral weight of the right peak decreases while that of the left peak increases. This redistribution of spectral weight between the two peaks is a hallmark of correlation-driven restructuring of the electronic spectrum. In the presence of Coulomb interactions, the eigenstates of the system are no longer simple single-particle states but rather correlated many-body states that incorporate the effects of electron-electron repulsion. The energy of these states depends on both the kinetic energy associated with interdot hopping and the potential energy arising from Coulomb repulsion. As $U$ increases, states with lower total Coulomb energy become relatively more favorable, and the system reorganizes its spectral weight to reflect this energetic preference. In the triangular geometry, certain many-body states correspond to charge configurations where electrons are distributed to minimize mutual repulsion while maintaining connectivity to the leads. These states can have energies that increase less rapidly with $U$ than other states, or in some cases, their energies relative to the Fermi level may even decrease due to the specific form of the interaction Hamiltonian in the triangular topology. This differential energy shift causes spectral peaks to move toward the Fermi level, enhancing transport when they enter the conducting window.

The non-monotonic behavior of the current is directly explained by the trajectory of the right spectral peak relative to the bias window centered at the Fermi level. At $U = 3t$, the right peak has shifted to a position near $\omega = -0.08t$, while the left peak resides at $\omega = -0.35t$. The right peak is now in close proximity to the Fermi level, and its spectral weight, although reduced compared to its value at $U = 2t$, still contributes significantly to the transport. As $U$ continues to increase and reaches $U = 3.5t$, the right peak arrives in the immediate vicinity of the Fermi level. At this point the left peak resides at $\omega = -0.31t$, having also shifted toward the Fermi level compared to its earlier position, but remaining well outside the narrow conducting window defined by the small applied bias. The transport current attains its maximum value at $U = 3.5t$ precisely because the antibonding spectral resonance is optimally aligned with the bias window. Physically, the antibonding spectral resonance, driven toward the Fermi level by the increasing Coulomb repulsion, sweeps through the narrow conducting window. The alignment of this resonance with the Fermi level maximizes the transmission probability and, consequently, the current.For $U$ larger than $3.5t$, the right peak continues to shift and eventually crosses the Fermi level, moving out of the conducting window on the opposite side. At the same time, the ongoing suppression of the spectral weight of this peak by the growing Coulomb repulsion further reduces its contribution to the current. The combined effect of the peak moving away from the optimal position and losing spectral weight produces the observed decrease in the transport current at large $U$. Although the left peak continues to approach the Fermi level slowly, but never enters the bias window, and its growing spectral weight cannot compensate for the loss of the right peak contribution.The physical mechanism driving this energy shift involves the interplay between Coulomb repulsion and the quantum interference inherent to the triangular geometry. In the triangular configuration, electrons can traverse between the two connected dots either directly through the interdot hopping or indirectly via the unconnected third dot. These two pathways interfere, and the relative phase between them depends on the energy spectrum of the system. Coulomb interactions modify the energy levels of the many-body states, thereby altering the interference conditions. In systems with quantum interference, the position of transmission resonances depends sensitively on the phase relationships between different transport pathways. Coulomb interactions introduce energy-dependent phase shifts that can tune these resonances into alignment with the Fermi level. This tuning effect is absent in linear geometries, where the lack of alternative pathways prevents the formation of interference-based resonances that can be manipulated by interactions.

To assess the robustness of this interpretation, we present in Figure3 the
transport current as a function of $U$ for several values of the dot-electrode
coupling strength $\Gamma$.

\begin{figure}[ht] 
	\centering 
	\includegraphics[width=0.5\textwidth]{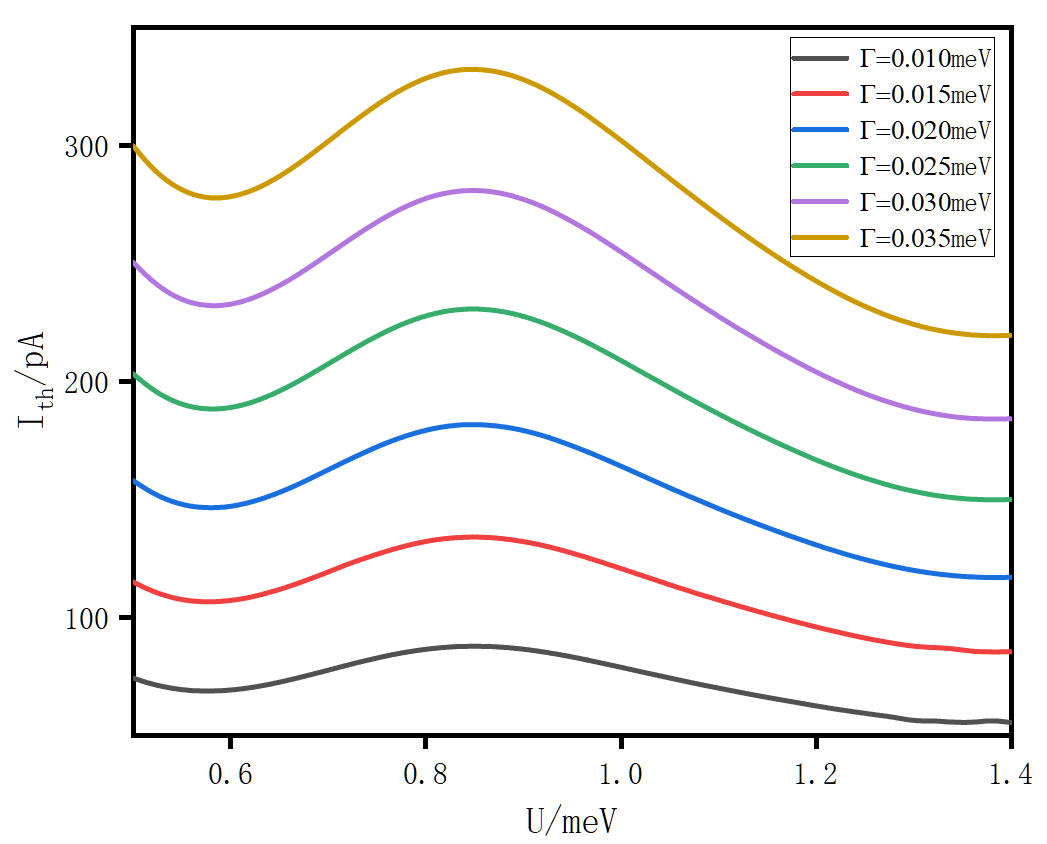} 
	\caption{(Transport current $I$ through the triangular triple quantum dot as a function of
    the on-site Coulomb interaction $U$, computed for several values of the lead-dot
    hybridization strength $\Gamma$.} 
	\label{fig:example-image} 
\end{figure}

Increasing $\Gamma$ systematically raises the magnitude of the transport current
across the entire range of $U$, which is physically expected, a larger $\Gamma$
broadens the quasiparticle resonances and increases the tunneling rate, thereby
enhancing the current for any given alignment between the spectral peak and the
transport window. The qualitative non-monotonic structure of
the $I$-$U$ curve is, however, preserved for all values of $\Gamma$ examined. The position
of the current maximum shifts only weakly with $\Gamma$, consistent with the fact
that the peak migration in energy is controlled by the many-body self-energy and
is therefore primarily a function of $U/t$ rather than of $\Gamma$. 

Reducing $\Gamma$ diminishes the current amplitude and sharpens the spectral
features, but again leaves the essential $U$-dependence unchanged. This
insensitivity of the current's qualitative behavior to $\Gamma$ is an important
consistency check: it confirms that the non-monotonic current is not an artifact
of a particular coupling regime but is a robust property of the triangular
topology and the associated chiral level renormalization. The spectral function
analysis of Fig.~2 and the coupling-strength sweep of Fig.~3 together provide
a consistent account of the anomalous current enhancement. The triangular
geometry generates chiral quasiparticle states whose energies are pulled toward
the Fermi level by the Coulomb interaction; as these states pass through the
transport window, the current rises and then falls. This behavior persists across
a wide range of $\Gamma$, confirming that it reflects the topology of the system rather than any fine-tuning of parameters.

To characterize the robustness of the anomalous current enhancement, we compute the transport current $I_{\mathrm{th}}$ as a function of both $U$ and $t$ over a wide range of parameters. The results are summarized in Figure4, which presents a three-dimensional plot of the threshold current $I_{\mathrm{th}}$ in the $t$-$U$ plane. The three-dimensional surface shows nontrivial structure. The yellow arrow in the figure traces a path of fixed $t$ along which $U$ increases, corresponding to the direction of current enhancement described in the previous section. Following the yellow arrow, the current rises from its value at small $U$, reaches a maximum, and then falls along the trajectory indicated by the blue arrow. The non-monotonic $U$-dependence of the current at fixed $t$ is thus consistently observed across the full range of $t$ values explored, confirming that the phenomenon is not limited to a specific choice of hopping amplitude but is a generic feature of the TTQD geometry.

\begin{figure}[ht] 
	\centering 
	\includegraphics[width=0.5\textwidth]{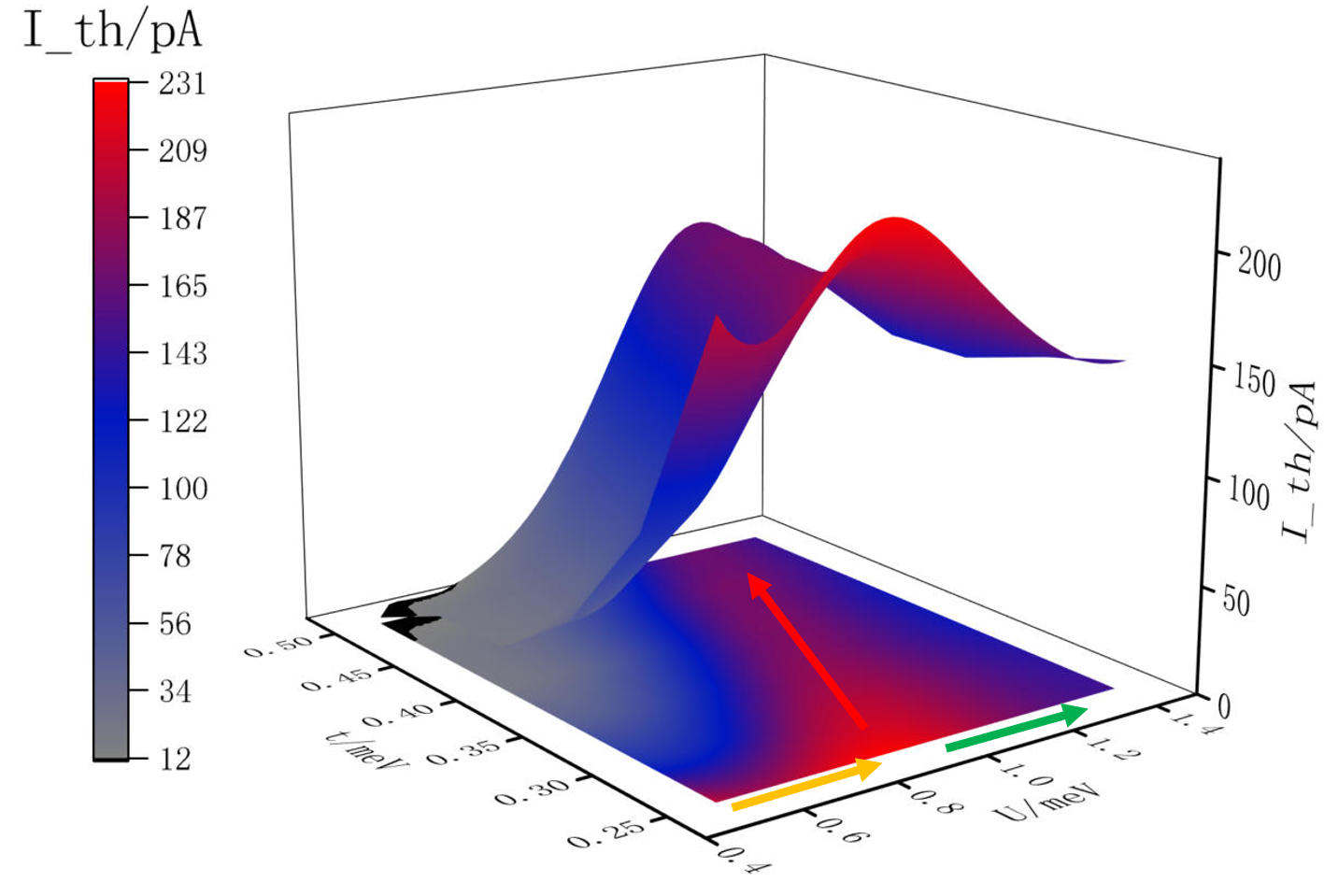} 
	\caption{Three-dimensional representation of transport current as a function of interdot hopping amplitude t and on-site Coulomb repulsion U for the triangular triple quantum dot system. Yellow arrows indicate regions where transport current increases with growing U, demonstrating the counterintuitive enhancement effect. Blue arrows mark regions where further increases in U suppress the current due to Coulomb blockade effects. Red arrows highlight the trend that larger values of t require correspondingly larger values of U to achieve maximum transport current, revealing that the phenomenon is governed by the dimensionless ratio U over t. The plot demonstrates that interaction-enhanced transport occurs in an intermediate regime where U and t are comparable in magnitude.} 
	\label{fig:example-image} 
\end{figure}

The red arrow in Figure4 reveals a systematic trend of particular physical significance, as $t$ increases, the value of $U$ at which the current reaches its maximum also increases. This means that the current peak shifts to larger $U$ as the hopping amplitude grows. The implication is that the anomalous current enhancement is a phenomenon governed by the interplay between $t$ and $U$ through their ratio, rather than by either parameter independently. This phenomenon has an important physical corollary. It means that the current enhancement is not a perturbative effect that vanishes at small $U/t$, nor is it a strong-coupling phenomenon that only appears at $U \gg t$. Instead, it occurs in an intermediate coupling regime where both kinetic and interaction energy scales are comparable, and where quantum interference in the triangular loop and Coulomb-driven renormalization of the chiral state energies cooperate to produce the anomalous transport signature. The linear geometry, lacking the loop structure, cannot support the closed-loop virtual processes that drive the orbital renormalization, and therefore shows no such enhancement regardless of the $U$ and $t$.

\section{Summary and Conclusions}

In this work, we have studied transport through a triangular triple quantum dot system. The central finding of this study is the counterintuitive non-monotonic dependence of the transport current on the on-site Coulomb interaction $U$. While a linear triple quantum dot under otherwise identical conditions shows a monotonically decreasing current as $U$ increases, the triangular system exhibits a clear current enhancement as $U$ is increased from moderate values, followed by a decrease at larger $U$. Spectral function analysis reveals that Coulomb repulsion drives certain many-body states toward the Fermi level, increasing their contribution to transport when they enter the conducting window defined by the bias voltage.  Finally, by mapping the transport current across a wide range of both $U$ and $t$, we have demonstrated that the current enhancement is governed not by either parameter alone but by the combined effect of $U$ and $t$. These results suggest that the triangular quantum dot geometry may serve as a useful experimental platform for engineering interaction-tunable transport responses, in which the current through the device can be increased rather than suppressed by strengthening the on-site Coulomb repulsion, provided the hopping amplitude and interaction strength are tuned to the appropriate intermediate-coupling regime.

The principle identified in this work, is not limited to the specific triangular three-dot geometry. More complex loop-containing networks, including tetrahedral four-dot clusters, ladder geometries, and two-dimensional arrays with embedded loops, are all expected to exhibit analogous interaction-driven spectral renormalization effects arising from the closed-loop virtual hopping processes identified here. As the network size and connectivity increase, the number of independent chiral channels grows, and the interplay between frustration, interaction, and interference is expected to give rise to increasingly rich transport phenomenology. The present work provides both the conceptual framework and the specific spectral mechanism needed to analyze and interpret these more complex cases, and it motivates a systematic study of how the topology of the dot network determines the sign and magnitude of the interaction-driven current correction across a broader class of correlated nanoscale conductors.

\section{Acknowledgements}  The support from the Natural Science Foundation of China (Grant Nos. 12274454, 11774418, 11374363, 11674317, 11974348, 11834014 and 21373191), the Strategic Priority Research Program of CAS (Grants No. XDB28000000 and No. XDB33000000), the Training Program of Major Research plan of NSFC (Grant No. 92165105).

\section{Data Availability}
The data that support the findings of this study are available from the corresponding author upon reasonable request.

\end{document}